# From Communication to Sensing : Recognizing and Counting Repetitive Motions with Wireless Backscattering

Ning Xiao*, Panlong Yang*, Yubo Yan*, Hao Zhou*, Xiang-Yang Li*, Haohua Du[†]
*University of Science and Technology of China, Hefei, China
[†]Department of Computer Science, Illinois Institute of Technology, Chicago, IL 60616 USA
xiaoning@mail.ustc.edu.cn, yanyub@gmail.com,
{plyang, kitewind, xiangyangli}@ustc.edu.cn, hdu4@hawk.iit.edu

———————————— • ————————————

**Abstract**—Recently several ground-breaking RF-based motion-recognition systems were proposed to detect and/or recognize macro/micro human movements. These systems often suffer from various interferences caused by multiple-users moving simultaneously, resulting in extremely low recognition accuracy. To tackle this challenge, we propose a novel system, called *Motion-Fi*, which marries battery-free wireless backscattering and device-free sensing. *Motion-Fi* is an accurate, interference tolerable motion-recognition system, which counts repetitive motions without using scenario-dependent templates or profiles and enables multi-users performing certain motions simultaneously because of the relatively short transmission range of backscattered signals. Although the repetitive motions are fairly well detectable through the backscattering signals in theory, in reality they get blended into various other system noises during the motion. Moreover, irregular motion patterns among users will lead to expensive computation cost for motion recognition.

We build a backscattering wireless platform to validate our design in various scenarios for over 6 months when different persons, distances and orientations are incorporated. In our experiments, the periodicity in motions could be recognized without any learning or training process, and the accuracy of counting such motions can be achieved within 5% count error. With little efforts in learning the patterns, our method could achieve 93.1% motion-recognition accuracy for a variety of motions. Moreover, by leveraging the periodicity of motions, the recognition accuracy could be further improved to nearly 100% with only 3 repetitions. Our experiments also show that the motions of multiple persons separating by around 2 meters cause little accuracy reduction in the counting process.

## 1 INTRODUCTION

### 1.1 Backgrounds and Motivation

**Device-free Interaction:** Emerging technologies in the wireless network have brought device-free interactions into reality. Due to the plausible features such as *non-invasive* installment [1]–[4], and *ubiquitous* deployment [5], [6], wireless signals are applied as sensors for human-computer interactions (HCI), behavior identification and movement measurements *et al.*. Although these applications and systems have achieved considerably high success, there are still some intrinsical limitations and deficiencies to conquer. On one hand, most of the existing systems are based on WiFi or RFID signals. These signals would be interfered by nearby wireless devices working in a similar spectrum, and suffer from the multi-path and fading effects as a consequence of surrounding layouts [7]–[9], which could be categorized as *inter-system interference*. On the other hand, with the increased number of Internet of Things (IoT) devices, dense deployment would result in severe interference among wireless devices, which could be categorized as *intra-system interference* [10]. Reducing the negative effect of inter-system and intra-system interferences is one of the grand challenges in these RF-based motion sensing/detection/recognition systems.

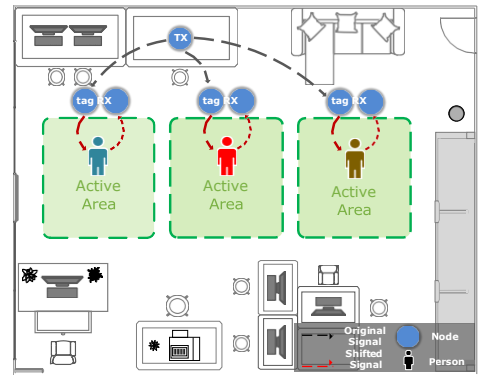

Fig. 1: Working Scenario of Motion-Fi.

**Battery-free Networking:** Recently, battery-free networking becomes prevalent under the persuasive need for large-scale and long-term IoT applications. With the advances of wireless technology, energy harvesting and backscatter communication have become two prevalent and major schemes for battery-free networking system. For energy harvesting, it relies on the transmission power and decaying law of propagation, which would be subjected to the charging distance and low energy transfer efficiency. While for backscatter communication, the information is transmitted with modulated Inspiringly, Passive-WiFi [1] is proposed, serving as an infrastructure-based powering system, which provides carrier with WiFi-like infrastructure and a high-frequency circuit to backscatter the modulated messages. This innovative design brings two foremost merits. First, the wireless APs can serve as a charging infrastructure to provide pervasive energy sources for devices. This vision is inspiring and impressive since the AP deployment is pervasive nowadays. Second, the modulated signal at the backscatter device could be transmitted at a rate up to 20Mbps due to the high-frequency circuit design. This is impressive, since RF transmission is the major source of energy consumption in such battery-free systems[1].

TABLE 1: Motion Types.

| Motion | Abbr. |
|--------|-------|
| Squats | SQ |
| Push-ups | PU |
| Sit-ups | SU |
| Leg-raise | LR |
| Step | ST |
| Stoop-down | SD |
| Dumbbell | DB |

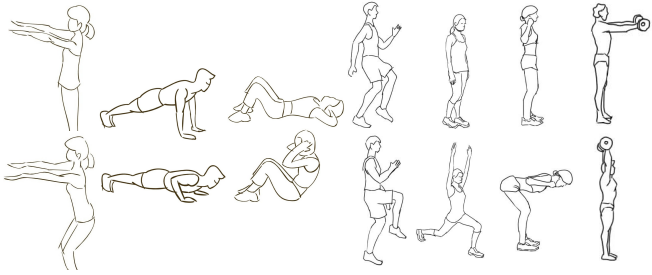

Fig. 2: Regular Motions.

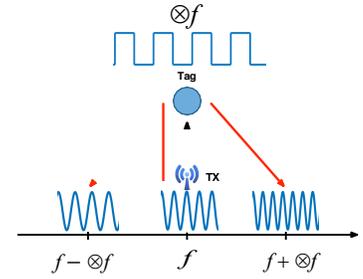

Fig. 3: Frequency Shift.

**Our Solution - Battery-free Motion Recognition:** In this paper, we propose *Motion-Fi*, a repetitive motion recognition system leveraging passive wireless backscattering paradigm. Here, we focus on two features of repetitive motions: periodicity (*counting*) and type (*classification*). As depicted in Fig. 1, it uses passive wireless backscattering devices and leverages wireless AP as their powering infrastructure [1]. We define 7 regular motions with abbreviations shown in Table 1 and the sketch of each motion is shown in Fig. 2. Compared with other battery-free systems such as RFID [11], it could be more pervasive due to the prevalent deployment of wireless APs. Moreover, signals could be collected in ambient ways, which differs from previous radar-like system [2], [3]. Compared with previous WiFi or RFID enabled device-free interactions, *Motion-Fi* would bring in the following irresistible advantages:

- In *Spatio Domain*, the working range of backscattering signals is relatively short, which reduces the *intra-system* interference significantly. This enables our system to support parallel motion recognition for multiple persons.
- In *Spectrum Domain*, our system design introduces a controllable frequency shift (*i.e.*, two mirror shifts in our current design) in the backscattered signal. By carefully selecting the frequency-shift, our system can avoid the *inter-system* spectrum occupancy, thus reducing the interferences.
- In *Deployment Domain*, our tags can be widely deployed just based on few signal sources. Moreover, the tags are cheap and small enough while consume little energy.

## 1.2 Challenges and Contributions

**Challenges:** Two challenges need to be formally addressed before realizing the aforementioned inspiring design vision with those favorable merits.

- Backscattering signals are weak but highly dynamic, especially when motions are incorporated. Previous motion recognition solutions often rely on building profiles as templates for recognizing different motions, which are not applicable in many scenarios. Such solutions have to revise the templates when the environment or setting changes.
- There exist irregular and user-dependent motion patterns, which often lead to incorrect motion recognition for different users. The effect of motion on the received signal varies in gender and body characteristics such as height and weight. Even when a person performs the same motion at different time, the motion's speed, direction, and angle may vary a lot. This ambiguity and diversity will lead to an exhaustive search of the entire signals, which will cause expensive computation costs. For a battery-free system, this inconvenience is intolerable.

To tackle these challenges, we demonstrate a prototype system, *Motion-Fi*, recognizing repetitive human motion frequency with passive wireless backscattering devices. *Motion-Fi* supports multi-user motion counting in various scenes and single-user motion recognition in a given environment. Compared with traditional WiFi-based gesture recognition systems, it supports motion recognition/counting in typical indoor scenarios without suffering from surrounding interference. Furthermore, we design a passive backscatter platform to validate our scheme. In our design, we first filter out the noise and observe the motion information through the baseband carrier. We propose an automatic matching scheme, where optimal patterns could be recognized for fre-quency identification and the irregularity in gestures could be effectively mitigated.

---

1. Maintaining frequency oscillator accuracy and ADC installation will cost most of the energy budget. High-speed backscattering scheme borrows the RF carrier from the infrastructure and achieve WiFi standard transmission by changing the resistance in circuit at nearly 20MHz.

Intrinsic properties of human motions are fully explored and an automatic signal segmentation scheme is proposed for counting the motions with high accuracy. Different from previous studies, our system works well even when motion patterns are not known in priori. Our extensive long-term experimental evaluation shows that our platform can effectively count the repetitive motions with an error rate less than 5%, across 26 different users, 7 motions, 4 orientations, and 5 distances for more than 6 months. Moreover, with just a little training effort via customized cubic-SVM, our system can accurately recognize different motion types with 93.1% recognition rate for 7 regular motions. Specifically, for periodical motions, the recognition rate could be further improved to nearly 100% after only 3 repetitions of the same motion.

The rest of this paper is organized as follows. Sec. 2 provides preliminary knowledge and presents the intrinsic principle on backscattering signal. After that, we introduce our system design and approaches in Sec. 3, and present the implementation details in Sec. 4. In Sec. 4, we also evaluate and validate our system with extensive experiments and analysis. Finally, we review the related work in Sec. 5 and conclude the paper in Sec. 6.

## 2 PRELIMINARIES AND SYSTEM MODEL

### 2.1 Preliminaries on Backscattering Signal

**Transmitter Side:** It provides a single-frequency tone as the plug-in device in passive WiFi design, and serves as a charging infrastructure. This vision is inspiring and impressive since passive WiFi deployment is considered to be pervasive in IoT industry. The tone signal can be simply represented as $\sin(2\pi f_c t)$.

**Backscattering Tag:** The backscattering tag is composed of an antenna and a micro-controller, which controls the SPDT RF switch network to generate backscatter signals. By changing the impedance of the antenna, the tag can switch its states between reflecting and non-reflecting. The scattered power of a passive tag is

$$P_{tag} = \frac{P_{tx} G_{tx} \Delta_\Gamma}{4\pi d^2},$$

where $P_{tx}$ and $G_{tx}$ is the transmission power and antenna gain of transmitter respectively, $d$ is the distance between WiFi AP and the tag, and $\Delta r$ is the differential Radar Cross Section (RDS) [2], [3] given by

$$\Delta_\Gamma = \frac{\lambda^2}{4\pi} G^2_{tag} \, |\Gamma^*_1 - \Gamma^*_2|,$$

where $\lambda$ is the wavelength of carrier, $G_{tag}$ is the antenna gain of tag and $\Gamma*$ is the conjugate reflection coefficient

$$\Gamma^* = \frac{Z^*_a - Z_c}{Z_a + Z_c},$$

where $Z_a = R_a + jX_a$ is the complex antenna impedance, and $Z_c = R_c + jX_c$ is the complex tag circuit impedance. By switching the impedance of the antenna would create additional narrow band signals whose frequency are shifted.

Let a square wave at a frequency of $\Delta_f$, which is generated by the micro-controller and used to control the impedance of the antenna. According to Fourier analysis, the first harmonic of a square wave is a sinusoid signal. Thus, we can simplify the process of square wave switching to sinusoid $\sin(2\pi\Delta_f t)$. Consequently, the process of backscattering can be represented by the product of the aforementioned two sinusoidal signals, which is given by

$$2\sin(2\pi f_c t)\sin(2\pi\Delta_f t) = $$
$$\cos(2\pi(f_c - \Delta_f)t) - \cos(2\pi(f_c + \Delta_f)t).$$

The frequency shifts of generated narrow band signals are $f_c - \Delta_f$ and $f_c + \Delta_f$. Fig. 3 illustrates the frequency shift.

**Receiver Side:** A receiver node tunes its center frequency to one of the shifted signal, which is $f_c - \Delta_f$ in our configuration. As described above, the received signal is generated by modulating the tone signal of transmitter and the sinusoid signal of backscattering tag in the air.

### 2.2 Frequency Shift

To study the signal path of our proposed system, we show the basic reflection model of gesture recognition with backscatter signal in Fig. 4a. The 'TX' node serves as a powering infrastructure, which sends a sinusoidal tone at a frequency of $f_c$ [1]. The 'Tag' node is a backscattering device, which switches the impedance of its antenna to *reflect* or *absorb* the tone signal. Noted that, in addition to reflect the tone signal transmitted by 'TX', there are two generated signals with frequency $f_c - \Delta_f$ and $f_c + \Delta_f$. A person noted as 'P' performs movements next to 'Tag'. A receiver noted as 'RX' tunes its center frequency to $f_c - \Delta_f$ to capture the generated signal. The sinusoid tone transmitted by 'TX' can directly traverse to 'RX', 'P' and 'Tag', with path $h_1$, $h_2$ and $h_3$ respectively. Then, 'P' reflects the signal to 'RX' and 'Tag' with path $h_7$ and $h_4$ respectively. 'Tag' reflects the sinusoidal signal at frequency $f_c$ to 'RX' with path $h_9$.

Meanwhile, the generated signals traverse from 'Tag' to 'P' and 'RX' with path $h_5$ and $h_8$. Then, 'P' reflects it to 'RX' with path $h_6$. As depicted in Fig. 4a, the traverse of original signal at the frequency $f_c$ is plotted in solid black lines, and the traverse of the shifted signal at the frequency $f_c - \Delta_f$ is plotted in dashed red lines. Since 'RX' receives signals at the frequency $f_c - \Delta_f$, we mainly study the signals plotted in dashed red lines.

### 2.3 Periodicity in Backscattering Signals

Since the channel is stable over a short period of time [12]–[14], it is believed that the changes of signal are mainly caused by human motion. As is shown in Fig. 4a, we classify a path as a dynamic one if the generated signal is affected by the person's motion when traversing across the path, *i.e.*, $h_5$ and $h_6$. Similarly, if the generated signal is not affected by the person's motion including both line-of-sight paths as well as the paths reflected from static objects, the path it traverses across is classified as a static one, *i.e.*, $h_8$. Here we believe that the signals after the human body (*i.e.*, $h_2$ & $h_4$.) is very weak compared to the signal that does not pass the human body (*i.e.*, $h_3$.) .

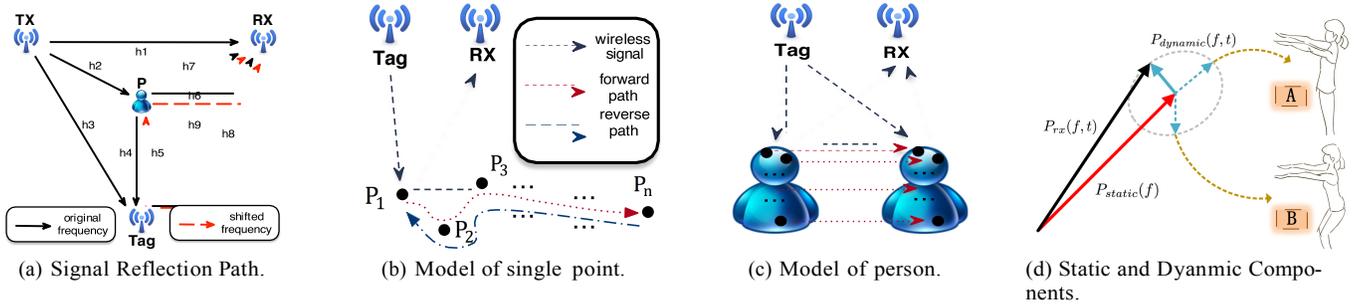

(a) Signal Reflection Path.   (b) Model of single point.   (c) Model of person.   (d) Static and Dyanmic Components.

Fig. 4: Principle on Backscattering based Motion Recognition.

Let the received energy by 'RX' is $P_{rx}$. Then, we can treat $P_{rx}$ as the sum of $P_{static}$ and $P_{dynamic}$ [4], that is:

$$P_{rx}(f, t) = P_{static}(f) + P_{dynamic}(f, t), \qquad (1)$$

where $P_{static}$ is the energy of signals received from static paths, and $P_{dynamic}$ denotes the reflected signal from human motion. According to the previous descriptions, we regard signals from path $h_8$ as static and signal from path $h_5$, $h_6$ as dynamic.

We can find that the dynamic signals is entirely generated by the movement of the human body. Further, we analyze the impact of human action on the dynamic part of recieved signal.

### 2.3.1 Model of single point

The simplest reflection model is one point, because there exists only one reflected path. If we consider the influence of both amplitude and phase, we can represent the dynamic singal reflected by the point use $a(f, t)e^{\frac{-j2\pi d(t)}{\lambda}}$, where $a(f, t)$ is the complex representation of amplitude and initial phase offset of the dynamic path, $d(t)$ denote the movement path of the object, and $e^{\frac{-j2\pi d(t)}{\lambda}}$ is the phase shift along the dynamic path length $d(t)$. As is shown in Fig. 4b, the points named $P_1$, $P_2$, $P_3$, ..., $P_n$ are the positions of the observation point at different sampling times, *i.e.*, $t_1$, $t_2$, $t_3$, ..., $t_n$. The signal from 'Tag' to point $P_i$ and then reflected to 'RX' is what we concerned as dynamic path.

As previously mentioned, we can get

$$P_{dynamic}(f, t_i) = a(f, t_i)e^{\frac{-j2\pi d(t_i)}{\lambda}}, i = 1, 2, 3, ..., n. \quad (2)$$

When we consider the periodic movement, the point will move from $P_1$ to $P_n$ and then back to $P_1$, as the forward path and reverse path depicted in Fig. 4b. Although the round-trip path can not be perfectly coincident, we can see the approximate symmetrical signal changes, which will be seen in our experimental part.

Moreover, if we denote

$$P_{dynamic}^k(f, t_i) = a^k(f, t_i)e^{\frac{-j2\pi d^k(t^k_i)}{\lambda}} \qquad (3)$$

as the dynamic signal of $k$-$th$ round-trip for a point periodic movement, we can further see cyclical changes in the signal due to the periodicity. Since the experience time of each cycle may be different, which will be shown on the received signal by stretching or compressing the signal pattern along the time axis, and we will describe how to solve this problem in the algorithm section.

### 2.3.2 Model of a person's motion

As for a person, we can divide a person's body into massive infitesimal segments, each fo these segments can be treat as a point as described above. That is, we can get the dynamci signal by

$$P_{dynamic}(f, t_i) = \sum_{segment_f \in Body} a_f(f, t_i)e^{\frac{-j2\pi d_f(t_i)}{\lambda}}, \qquad (4)$$

$$i = 1, 2, 3, ..., n.$$

Since the motion of each segment can be considered to be periodic, the sum function of the periodic function also has a periodic signal that can be used to obtain periodic motion of the human body.

We demonstrate this principle in Fig. 4d. So, if we perform periodical motions, the signal of $P_{dynamic}(f, t)$ also changes periodically, thus the energy of $P_{rx}(f, t)$ is periodical.

## 3 SYSTEM DESIGN

### 3.1 Design Overview

Our *Motion-Fi* design can recognize/count repetitive motions regardless of the motion, even when multiple users are performing movements at the same time. To achieve this goal, the following three challenges must be properly addressed in the system design. First, how to process the received signals to eliminate the impact of noise? Second, how to design the segmentation scheme to count motions despite of their types? Third, how to classify the type of motions with lower cost in a given environment? We address each of these questions in the following subsections.

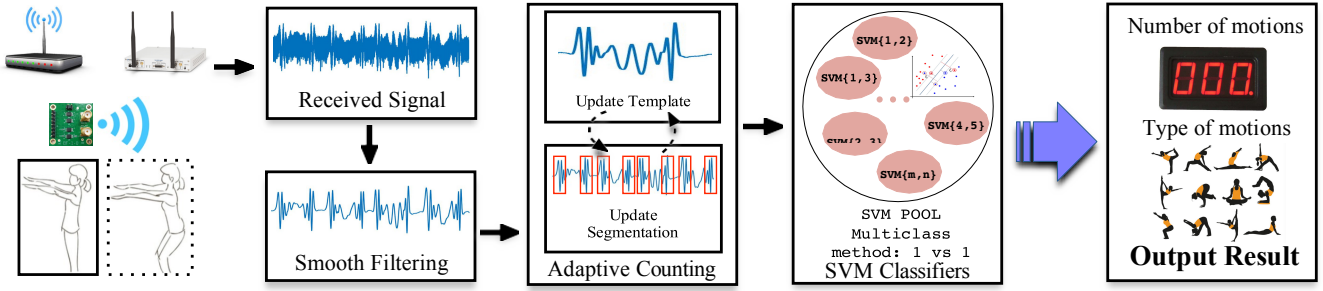

Fig. 5: System Overview.

The system overview of *Motion-Fi* is illustrated in Fig. 5. There are mainly three modules in *Motion-Fi*, which are *smoothing filter*, *adaptive counter* and *SVM classifiers*. When the frequency shifted backscattering signal is received, the energy of signal is computed according to Sec. 3.2. Then, an LPF filter is applied to the received signal to smooth the amplitude. The adaptive counting module is used to count the repetitive motions without using scenario-dependent templates or profiles. After that, an enhanced SVM classifier is used to recognize the motions. Finally, *Motion-Fi* outputs the counting number and type of motions.

## 3.2 Preprocessing of the Signal

As is known, $I(t)$ and $Q(t)$ are orthogonal signals in wireless communications. They can be represented as

$$I(t) = A(t)A_0\cos(2\pi ft) + N_I(t),$$
$$Q(t) = A(t)A_0\sin(2\pi ft) + N_Q(t),$$
(5)

where $A_0$ and $f$ are the amplitude and frequency of the sinusoidal signal. $N(t)$ is the white noise. $A(t)$ denotes the influence coefficient[2] of motions, which is equal to 1 when there are no motions.

Then the energy of received signal is $E(t) = (I(t)^2 + Q(t)^2)$, and replace $I(t)$ and $Q(t)$ with Eq. (5). After that, $E(t)$ could be represented as

$$E(t) = A(t)A_0 + N^\dagger(t),$$
(6)

where $N^\dagger(t)$ is a term relevant to the white noise. Compared with human movements, $N^\dagger(t)$ can be treated as high frequency noise, which could be filtered out with LPF (Low Pass Filter). After applying LPF to $E(t)$, we have:

$$E^\dagger(t) = A(t)A_0.$$
(7)

Finally, the filtered signal could reflect human movements approximately, since $A_0$ is constant in this scenario.

---

2. In our system, the effect of human motions on signals is reflected in the change of signal amplitude, which is equivalent to multiplying a coefficient in the original amplitude.

## 3.3 Motion Segmentation

We formulate an optimization problem that jointly recovers the morphology of the motions and the segmentation [5]. We adjust the algorithm to fit different motions, whose cyclical features are weaker than those of heart beats. The intuition underlying this optimization is that successive human motions should have similar morphology. Since each motion lasts different time duration, we can define the similarity of the two motions by stretching or compressing the motion signal to an equal length. So the goal of the algorithm is to find an optimized segmentation which makes the difference between each segmentation as small as possible, while accounting for the fact that we do not know a priori the shape of a motion.

Let $\underline{x} = (x_1, x_2, ..., x_n)$ denote a sequence of length $n$. A segmentation $\mathbf{S} = \{\underline{s_1}, \underline{s_2}, ...\}$ of $\underline{x}$ is a partition that divides $\underline{x}$ into non-overlapping continuous subsequences (segments), where each segment consists of points and $|\underline{x}| = \sum_{\underline{s} \in \mathbf{S}} |\underline{s_i}|$.

In order to identify each motion cycle, our idea is to find a segmentation with segments most similar to each other, *i.e.*, to minimize the variation across segments. Since statistical distance is only defined for scalars or vectors with the same dimension, we extend the definition of distance for vectors with different lengths as follows:

*Definition 1:* The distance between vectors $\underline{a_1}$ and $\underline{a_2}$ is:

$$Dist(\underline{a_1}, \underline{a_2}) = dtw(\underline{a_1}, \underline{a_2}),$$

where $dtw(\underline{a_1}, \underline{a_2})$ is dynamic time warping function to measure the similarity between two temporal sequences. It stretches two vectors, $\underline{a_1}$ and $\underline{a_2}$, onto a common length $L$ ($\max(|\underline{a_1}|, |\underline{a_2}|) \leq L \leq |\underline{a_1}| + |\underline{a_2}|$), making the sum of the Euclidean distances between corresponding points the smallest. To stretch the inputs, $dtw$ will repeat each element of $\underline{a_1}$ and $\underline{a_2}$ as many times as necessary.

Following that, we define the distance between segmentation $\mathbf{S}$ and motion template $\underline{\xi}$[3]:

*Definition 2:* The distance between segments $\mathbf{S} =$

---

3. $\underline{\xi}$ is a vector represents the central tendency of all the segments, i.e., a template for the motion shape (or morphology). We denote m=$|\underline{\xi}|$ in the following content.

$\{s\_1, s\_2, ...\}$ and a template $\boldsymbol{\xi}$ is:

$$Dist(\mathbf{S}, \boldsymbol{\xi}) = \sum_{s_i \in \mathbf{S}} Dist(\underline{s_i}, \boldsymbol{\xi}) = \sum_{s_i \in \mathbf{S}} dtw(\underline{s_i}, \boldsymbol{\xi}).$$

The goal of our algorithm is to find the optimal segmentation $\mathbf{S}*$ and template $\boldsymbol{\xi}*$ that minimizes $Dist(\mathbf{S}, \boldsymbol{\xi})$. Since both segmentation $\mathbf{S}$ and template $\boldsymbol{\xi}$ are unknown for us, we can rewrite it as the following joint optimization problem:

$$(\mathbf{S}*, \boldsymbol{\xi}*) = \arg\min_{\mathbf{S}, \boldsymbol{\xi}} Dist(\mathbf{S}, \boldsymbol{\xi})$$

$$= \arg\min_{\mathbf{S}, \boldsymbol{\xi}} \sum_{s_i \in \mathbf{S}} dtw(\underline{s_i}, \boldsymbol{\xi}),$$

$$s.t. : t_{min} \leq \frac{|\underline{s_i}|}{C} \leq t_{max}, \underline{s_i} \in \mathbf{S},$$

where $t_{min}$ and $t_{max}$ are constraints on the length of each motion cycle[4], and $C$ denotes the sample rate. It tries to find the optimal segmentation $\mathbf{S}$ and template (*i.e.*, morphology) $\boldsymbol{\xi}$ that minimize the sum of the distance between segments and template. This optimization problem is difficult as it involves both combinatorial optimization over $\mathbf{S}$ and numerical optimization over $\boldsymbol{\xi}$. Exhaustive search of all possible segmentations will lead to exponential complexity.

### 3.4 Iterative Segmentation

Instead of estimating the segmentation $\mathbf{S}$ and the template $\boldsymbol{\xi}$ simultaneously, our algorithm alternates between updating the segmentation and template. During each iteration, our algorithm updates the segmentation given the current template, then updates the template given the new segmentation. For each of these two sub-problems, our algorithm can obtain the global optimal with linear time complexity.

**Update Segmentation:** In the $i$-th iteration, segmentation $\mathbf{S}^{i+1}$ is updated given template $\boldsymbol{\xi}^i$ as

$$\mathbf{S}^{i+1} = \arg\min_{\mathbf{S}} Dist(\mathbf{S}, \boldsymbol{\xi}^i). \quad (8)$$

Though the number of possible segmentations grows exponentially with the length of $x$, the above optimization problem can be solved efficiently using dynamic programming [15]. The recursive relationship for the dynamic program is as follows: if $D_l$ denotes the minimal cost[5] of segmenting sequence $\underline{x}_{1:l}$, then we have

$$D_l = \min_{\tau} \{D_\tau + Dist(\underline{x}_{\tau+1:l}, \boldsymbol{\xi})\}, \quad (9)$$

where $(l = \{t_{min}, t_{max}\}$ and $l - C * t_{max} \leq \tau_{l,2} \leq l - C * t_{min}$ specifies possible choices of $\tau$ considering the segment length. According to Eq. (9), the time complexity of the dynamic programing is $O(C(t_{max} - t_{min})n)$ and the global optimum could be guaranteed.

4. $t_{min}$ and $t_{max}$ capture the fact that human motion cannot be too short or too long.

5. The minimal cost of sequence $\underline{x}_{1:l}$ is the minimum of $Dist(\mathbf{S}, \boldsymbol{\xi})$ in this sequence.

---

**algorithm 1** Motion Segmentation

---

**Input:** sequence $\underline{x}$, time constraint $(l = \{t_{min}, t_{max}\}$
**Output:** segments $\mathbf{S}$, template $\boldsymbol{\xi}$
1: initialize $\boldsymbol{\xi}^0$ as zero vector ;
2: $m \leftarrow \frac{1}{2} * (t_{max} + t_{min}) C_k$ ;
3: $i \leftarrow 0$ ; //number of iterations
4: $Similarity \leftarrow Inf$ ;
5: **while** $(Similarity \geq Threshold)$ **do**
6:  $\mathbf{S}^{i+1} \leftarrow UpdateSegmentation(\underline{x}, \boldsymbol{\xi}^i)$ ;
7:  $\boldsymbol{\xi}^{i+1} \leftarrow UpdateTemplate(\underline{x}, \mathbf{S}^i)$ ;
8:  $i \leftarrow i + 1$ ;
9:  $Similarity \leftarrow Dist(\boldsymbol{\xi}^{i+1}, \boldsymbol{\xi}^i)$ ;
10: **end while**
11: **return** $\mathbf{S}^i$ *and* $\boldsymbol{\xi}^i$

12:
13: **function** UPDATESEGMENTATION($\underline{x}, \boldsymbol{\xi}$)
14:  $\mathbf{S}_0 \leftarrow \varnothing$ ;
15:  $D_0 \leftarrow 0$ ;
16:  **for** $t = 1 \rightarrow length(\underline{x})$ **do**
17:   $\tau^* \leftarrow \arg\min_{\tau \in t_{l,2}} \{D_\tau + Dist(x_{\tau+1:t}, \boldsymbol{\xi})\}$ ;
18:   $D_t \leftarrow D_{\tau*} + Dist(\underline{x}_{\tau+1:t}, \boldsymbol{\xi})$ ;
19:   $\mathbf{S}_t \leftarrow \mathbf{S}_{\tau*} \cup \{\underline{x}_{\tau*+1:t}\}$ ;
20:  **end for**
21:  **return** $\mathbf{S}_n$
22: **end function**
23:
24: **function** UPDATETEMPLATE($\underline{x}, \mathbf{S}$)
25:  $n \leftarrow length(\underline{x})$ ;
26:  $\boldsymbol{\xi} \leftarrow \frac{1}{n} \sum_{s_i \in \mathbf{S}} |\underline{s_i}| \omega(\underline{s_i}, m)$ ;
27: **end function**

---

**Update template:** In the $i$-th iteration, template $\boldsymbol{\xi}^{i+1}$ is updated given segmentation $\mathbf{S}^i$ as

$$\boldsymbol{\xi}^{i+1} = \arg\min_{\boldsymbol{\xi}} \sum_{s_j \in \mathbf{S}^i} Dist(\underline{s_j}, \boldsymbol{\xi}). \quad (10)$$

Obviously, the above optimization is a weighted least squares problem and it is impossible to meet the above optimization goals $\boldsymbol{\xi}^{i+1}$ by using brute-force search. Here we present an analogy method to get a feasible solution.

Consider another similar problem: given an ar-ray $\underline{a}$, find a number $x$ such that the sum of the distance between $x$ and each element in ar-ray $\underline{a}$ is minimized. If $y$ denotes the sum of distance, we can get the expression as follows easily: $y = \sum_{i=1}^{|a|} (a[i] - x)^2 = |\underline{a}|x^2 - 2C_1x + C_2$, where $C_1$ and $C_2$ are two constants, $C_1 = \sum_{i=1}^{|a|} a[i]$, $C_2 = \sum_{i=1}^{|a|} a[i]^2$. Obviously, this is a quadratic function optimization problem, and it is easy to get the solution that makes y minimal: $x* = \frac{(-2*C_1)}{2*n} = \frac{1}{|a|} \sum_{i}^{|a|} a[i] = mean(\underline{a})$.

From the above equation we can see that the optimal solution $x*$ is the average of the array $\underline{a}$. Similarly, we map the problem to Eq. (10), the array element $a[i]$ becomes the subsequence $\underline{s_i}$, $x$ becomes the template vector $\boldsymbol{\xi}$, the distance between two values becomes the distance of two vectors, in the end, the solution $x*$ is

TABLE 2: Recognition Accuracy for Candidate Methods.

| Classification Method | Accuracy |
|---|---|
| MLP | 86.2% |
| Ensemble | 91.4% |
| Decision Tree | 88.4% |
| Discriminant Analysis | 68.3% |
| Nearest Neighbor Classifier | 83.8% |
| Cubic Kernel SVM | **93.1%** |
| Linear Kernel SVM | 86.9% |
| Gaussian Kernel SVM | 90.2% |
| Quadratic Kernel SVM | 92.3% |

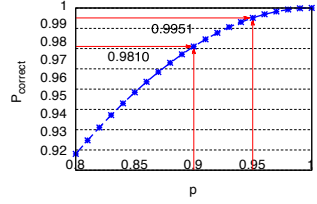

Fig. 6: Voting Strategy.

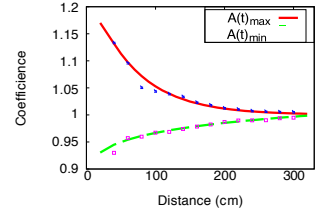

Fig. 7: Impact of Distance.

mapped to the weighted average of the subsequences, and a feasible solution is given as follows:

$$\boldsymbol{\xi}^{i+1} = \frac{\sum_{s_j \in \mathbf{S}^I} |s_j| \omega(s_j, m)}{\sum_{s_j \in \mathbf{S}^I} |s_j|} = \frac{1}{|x|} \sum_{s_j \in \mathbf{S}^I} |s_j| \omega(s_j, m), \quad (11)$$

where the weight of each subsequence is the proportion of the length of the sequence to the entire sequence, $\omega(s_j, m)$ is linear warping of $s_j$ into length $m$, which is realized through a cubic spline interpolation [16].

The algorithm pseudo-code is shown in Algorithm 1. The stopping condition represents the convergence state. The difference between two conjunctive iterations are inspected as the criterion of convergence. For practical consideration, we set a feasible threshold, where system efficiency and accuracy can be achieved.

### 3.5 Motion Classification

The classification of motions is based on the cubic SVM method. After segmenting the signal, we carry out the signal normalization processing. For a sequence $\underline{x} = (x_1, x_2, ..., x_n)$, the processing method is:

$$x_i^t = \frac{x_i - \min(\underline{x})}{\max(\underline{x}) - \min(\underline{x})}, i = 1, 2, 3, ..., n, \quad (12)$$

where $\max(\underline{x})$ and $\min(\underline{x})$ denote the maximum and minimum of $\underline{x}$.

After normalization, we extract the following features [6] of new sequence $\underline{x}^t$. First of all, the mean and standard deviation values are incorporated, which are given by $\mu = \frac{1}{|x|} \sum_{i=1}^{|x|} x_i^t$, and $\sigma = \sqrt{\frac{\sum_{i=1}^{|x|} (x_i^t - \mu)^2}{|x|}}$. Second, the maximum and minimum values of $\underline{x}^t$ are considered. Typically, the 3 quantiles ($p = 0.25, 0.5, 0.75$), skewness ($Skew(x^t) = \frac{E(x-\mu)^3}{\sigma^3}$), kurtosis ($Kurtosis(x^t) = \frac{E(x-\mu)^4}{\sigma^4}$) and $\theta = \sqrt{\frac{\sum_{i=1}^{|x|} x_i^{t^2}}{|x|}}$ are incorporated as features of our SVM model.

After comparing different classification methods such as decision trees, discriminant analysis, nearest neighbor classifiers (KNN), we find that SVM is a stable and effective one for our work. Moreover, we have tried many kernel functions as well, including *polynomial* kernel and *Gaussian* kernel. As Table 2 shows, the cubic kernel is the best one. Consequently, the kernel function we use is

$$K(\underline{v_i}, \underline{v_j}) = (1 + \underline{v_i}^T \underline{v_j})^3, \quad (13)$$

where $\underline{v_i}$ and $\underline{v_j}$ represent the feature vectors. Besides, we use a one-vs-one multi-classification approach for our model.

Moreover, since our counting algorithm has segmented the experimental data, given a series of $k$ motions, we can provide $k$ segmentations to the classifier and will get $k$ results in return. Thus we can use voting method to get a more accurate result. Owing to the judgment for each segmentation is independent, we can denote the probability of judging correctly once as $p$, and for three (here we denote $k = 3$) consecutive segments (they are the same motion) the correct probability is

$$\begin{aligned} P_{correct} &= p^3 + \binom{3}{2} \cdot p^2(1-p) + \binom{3}{1} \cdot p(1-p)^2, \\ &= p + p^2 - p^3. \end{aligned} \quad (14)$$

Fig. 6 shows the relationship between $P_{correct}$ and $p$. As depicted in Fig. 6, when $p = 0.9$, our accuracy can further be improved to 98.10%, when $p$ increased to 0.95, the accuracy will be raised to 99.51% using three repetitions!

## 4 IMPLEMENTATION AND EVALUATION

### 4.1 Implementation

We implement a prototype of *Motion-Fi* using COTS components for signal backscatter and an USRP-RIO for signal receiving and processing. The prototype of *Motion-Fi* is built with our customized design of passive tag, which backscatters the signal of plug-in device to the receiver. Fig. 8a shows our prototype of a passive tag, whose main components are SPDTs, antenna interfaces, pins and resistances, the size of which is about $3.4 \times 3.4 \ cm^2$. In order to enable our tag adapt to high-frequency signal, we choose the HMC190BMS8 SPDT. An additional FPGA is used to control the backscatter state of the passive tag. The reflected signal is received by a NI USRP-2953R, and then fed to our proposed scheme to recognize repetitive motions. Another USRP serves as a plug-in device to transmit signal in the environment. Our equipments are shown in Fig. 8b.

Our software platform is built upon LabVIEW, where coding program could be put into our aforementioned hardware platform. First of all, we need a carrier centering at 2GHz. Our FPGA program built with LabView is applied to produce a square wave at a frequency shift. It is worth noting that, in practical design, a 'perfect' square wave is not available[6]. Thus we use the sine wave to approximate. From Fourier analysis, a square wave can be written as:

$$Square(\Delta f * t) = \frac{4}{\pi} * \sum_{n=1,3,5,...}^{\infty} \frac{1}{n} \sin(2\pi n \Delta f * t) \quad (15)$$

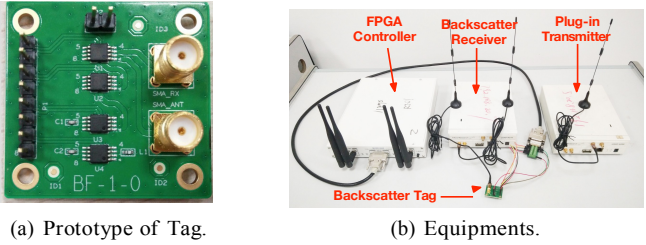

(a) Prototype of Tag.　　　　(b) Equipments.

Fig. 8: Demonstration Equipments.

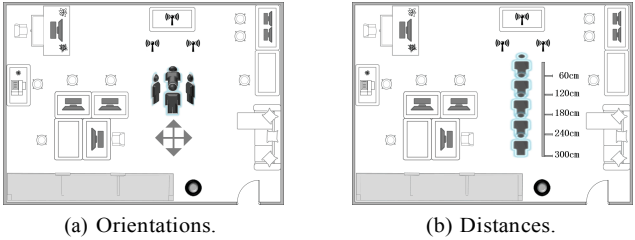

(a) Orientations.　　　　(b) Distances.

Fig. 9: Configurations of Persons on Orientation and Distance.

Here the first harmonic is a sinusoidal signal at the desired frequency $\Delta f$. Note that the power in each of these harmonic scales as $\frac{1}{n^2}$. So the third and the fifth harmonic are around 9.5dB and 14dB lower than the first harmonic. Thus, we can approximate a square wave as just the sinusoidal signal, $\frac{4}{\pi}\sin(2\pi\Delta f * t)$.

### 4.2 Experiment Settings

**Deployment and Layout:** We analyze the performance of our system in a typical office environment, which covers a $6.3 \times 5.7$ $m^2$ area, consisting of some bookcases and office furniture, including desks, chairs and computers. One USRP serves as a plug-in device for single-frequency tone signal transmission and another one serves as a receiver. The transmitter is placed in a desktop as shown in Fig. 1, and the receiver and tag pair are placed in the front of the gym area. When multiple persons performing motions, each person does it in an area $2m^2$ in front of one tag (as shown in Fig. 1).

Fig. 9 depicts the orientations and distances of the persons involved in evaluations. Basically, there are four typical directions(Front, Rear, Left and Right) to evaluate the impact of orientations. Each orientation denotes the relative position between a person and tag. For example, Front means the tag is in front of the person. Meanwhile, we test five distances from 0.6m to 3$m$. The maximum distance is set to 3$m$, because the backscattering signal is too weak to be used for recognition at this distance according to our evaluations (Fig. 7).

---

6. The transition between minimum to maximum is instantaneous for an ideal square wave, but this is not realizable in physical systems.

Fig. 10 shows two typical deployment of **Tag**, **TX** and **RX**. As depicted in Fig. 10a, the **Tag** is away from the very near **TX** and **RX** pair. It works as if **Tag** is a client, which is affected by the server side **TX** and **RX** pair. We use this evaluation to show the impact of **Tag** distance. The near deployment of **TX** and **RX** is similar to RFID reader. While for Fig. 10b, the difference is **TX** and **RX** are deployed seperately with each other. This mode could be utilized when WiFi infrastructure and receiving devices are working independently and distributedly.

Moreover, we find that the strength of the received signal is only related to the path **Tx** > **Tag**−> **Rx**, and has nothing to do with the path between **Tx** and **RX**. Consequently, we fix the positions of **Tx** and **Rx**, observe the change of received signal when **Tag** moved between them, as is shown in Fig. 10c. We can see that the strength of the received signal increases when the tag get close to either **Tx** or **Rx** and it achieves the minimum at the midpoint of the connection between **Tx** and **Rx**. This is well understood since the signal strength is proportional to $\frac{1}{d^* d}$, and the experimental result is consistent with this mathematical meaning.

Finally, we treat **Tx** as a infrastructure that can not moved easily. So, to in order to obtain a stronger received signal and enable **Tx** to cover further distance, we place the **Tag** and **Rx** together as depicted in Fig. 1.

**Volunteers and Concurrent Motions:** We recruited 26 volunteers, including 16 males and 10 females for testing over 6 months. They vary in age (18-45 years old), stature (155-188$cm$) and weight (45-90$kg$). During the experiments, they wore their daily attire with different fabrics and they performed motions at the specified location within the scope of the $2m^2$ square area. The variance in experimental environment and the presence of other users had a negligible impact on the results, because the backscatter signal is relatively weak, where only nearby motions can cause discernible volatility on the receiver signal strength. In our scenario, up to 3 persons could perform motions concurrently. We then show all experimental results in the following benchmark tests.

### 4.3 Long-term Experimental Results

We set up a gym area to evaluate the performance of *Motion-Fi* for more than half a year. Volunteers perform daily exercise in the gym, and our equipments record the data at the same time. We record the meta-data for each exercise, including name, time, motion, number, *etc.*. We totally record thousands test data. Each test contains one of the 7 motions as listed in Table 1. One motion is repeated from 20 to 80 times during each test. We evaluate our proposed scheme on each test data, and examine the error ratio of counting for motions. The error ratio of counting is defined as

$$error\ ratio = \frac{N_{est} - N_{truth}}{N_{truth}} \times 100\%,$$

where $N_{est}$ is the number of estimated motion counting, and $N_{truth}$ is the number of recorded motion counting. Note error ratio in one test could be negative. The evaluation results are averaged on the test records.

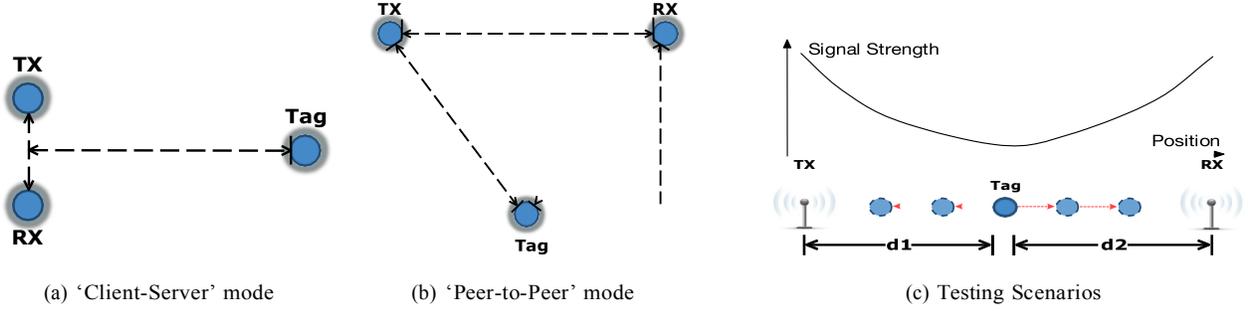

(a) 'Client-Server' mode     (b) 'Peer-to-Peer' mode     (c) Testing Scenarios

Fig. 10: Deployment of Tags, Transmitter and Receiver.

**Basic experimental observation:**

We first observe the received signal when one person performs different motions, different persons perform the same motions and one person perform the same action at different distances, the result of which are shown in Fig. 11.

Form Fig. 11a, we can see the basic periodicity in the time series and for each fragment we can also see the symmetry when different persons perform the same motion (SQ), which is consistent with what we described in Sec. 2.3. Besides, we can find the difference (*i.e.*, different waveforms, different cycle times and different amplitude changes) among these signals, because there exist different habits when different people perform the same action, such as the speed of motion, the degree of leg bending and body differences (*i.e.*, height and weight).

Fig. 11b show the result when one person performs 3 kinds of motions (SQ, PU and SU), from which we can also see the periodicity and symmetry. However, the waveforms of them vary a lot since different motions have distinctly different modes, which can be used to distinguish different motions.

Fig. 11c depicts the distance effect when one person perform squats (SQ) at 3 distances (20cm, 60cm and 100cm). Since the signal is attenuated square inversely with the increase in distance, we can see the dynamic part of received signal decay quickly. This enable our system to support multiple persons perform simultaneously.

**Performance of single person:** We evaluate the error ratio of motion counting for various motions, persons, orientations and distances when only one person per-forms motions. To do this, we place a transmitter, a receiver and a tag in the room as shown in Fig. 9, the transmission power is set to 10dBm.

We firstly study the impact of distance on the strength of received signal. The distance from tag to person varies from 40cm to 300cm with step of 20cm. Specifi- cally, a volunteer performs 100 motions at each location apart from tag 20, 40, , 300cm respectively. Fig. 7 shows the changes of $A(t)$ with the increase in distance, where $A(t)$ denotes the coefficient on received signal with human motions, $A_0$ denotes the amplitude without human motions (Eq. (7)), $A(t)_{max} = \max(E'(t))/A_0$ is the maximum value of $A(t)$, and $A(t)_{min} = \min(E'(t))/A_0$ is the minimum value of $A(t)$. By showing the difference of $A(t)_{max}$, $A(t)_{min}$ respectively, we conclude that the effect of human motion on received signal intensity is less than 2% when the distance is beyond 2 meters. This is a favorable property for the application such as exercising in a gym as this observation enables motion counting when a lot of people doing exercise simultaneously with a small seperation distance among them.

We choose one volunteer to perform all the seven motions, and examine the error ratio of counting for each of the motions. The results are shown in Fig. 12a. It can be seen that the error ratio of counting are different among motions, from 0 to 2%. Noted that the error ratio of Push-ups (PU), Stoop-down (SD) and Dumbbell (DB) are relative small, while the error ratio of Squats (SQ), Sit-ups (SU), Leg-raise (LR) and Step (ST) are larger. The reason is that when a person performs these type of motions, part of the limb's range of activities is relative large, which lead to larger variances of received signal.

To study the changes of error ratio when different persons perform the same motion, we arrange for 6 volunteers labeled with P1 to P6 to perform squats one by one. The results are shown in Fig. 12b. In general, the average of the error ratio is no more than 2%. However, there do exist difference among different people, *i.e.*, when different people do squat motion, their relative movement of arms is not exactly the same and some people can not maintain the consistency of motions during our test.

To study the impact of orientations when people do motions, we let a volunteer do squats in 4 orientations as shown in Fig. 9a. The evaluation results are shown in Fig. 12c. It can be seen that the error ratio of Front and Rear are lager than that of Left and Right. The reason is that when people do motions with orientations of Front and Rear, the cross section of their body is larger than that of Left and Right. Thus, inconsistency of motions with orientations of Front and Rear will cause large fluctuations in backscattered signals.

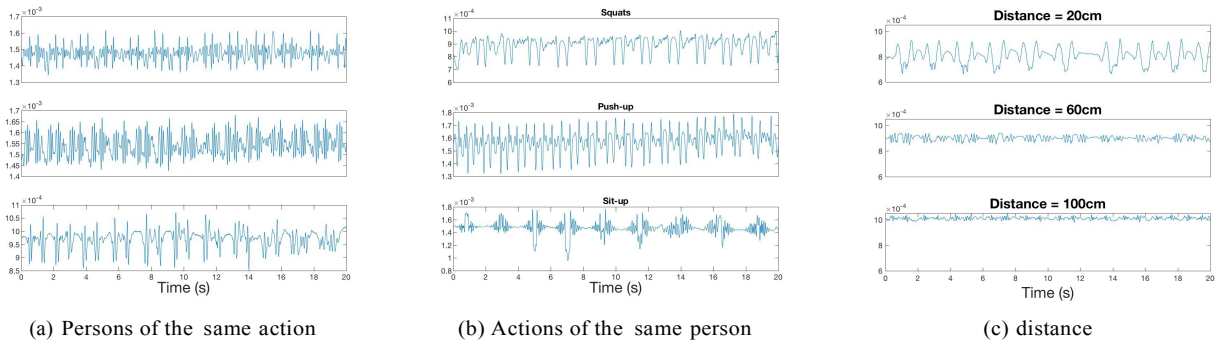

(a) Persons of the same action      (b) Actions of the same person      (c) distance

Fig. 11: Diversities in persons and differences in actions

Finally, we study the impact of distance between person and tag on the counting error ratio. A volunteer performs squats at 5 distances ranging from 60*cm* to 300*cm* as shown in Fig. 9b. The error ratio at different distances are shown in Fig. 12d. We can see that when the distance is ⩾100*cm*, the error ratio increases because the strength of backscattering signal decreases with distance. Noted that, when the distance is ⩽ 100*cm*, *e.g.* 60*cm* in our tests, the error ratio increases as distance decreases. Although the strength of backscattering signal increases when distance decreases, the interference caused by the irregular movement of limbs also increases. Thus, the distance ∼ 1*m* is ideal for our applications.

**Performance of multiple persons:** We then evaluate the performance of *Motion-Fi* when multiple persons perform motions simultaneously. We deploy three sets of equipments in our laboratory as shown in Fig. 1. Specifically, we deploy one transmitter to transmit the tone signal. For each person, a pair of tag and receiver is used to recognize his/her motion. During our evaluation, three persons perform motions in the test area simultaneously. Each person stands before the tag at a distance about 1*m*. Three persons are labeled with 'Left', 'Middle' and 'Right' respectively. We examine the error ratio of motion counting under various conditions.

We firstly study the impact of transmission power of the tone signal. To do this, we configure the transmission power to 0, 10 and 20dBm respectively. Three persons perform squats in front of the tag at a distance of 1*m* simultaneously and the separation between them is 2*m*. Fig. 13a shows the error ratio of motion counting. It can be seen that there is a trade-off between transmission power and error ratio. In our experiment configuration, a transmission power of 10dBm can achieve the lowest error ratio. Noted that when transmission power is too weak, for example 0dBm in our case, the backscattered signal becomes too weak to discern. However, when transmission power is too strong, the interference from adjacent person becomes more serious. So, we configure the power to 10dBm in the following tests.

We then examine the impact of separation distance between users. Again, three volunteers perform squats simultaneously. The distance between tag and person is about 1 *m*. We change the separation between persons from 150*cm* to 250*cm* at a interval of 50*cm*. The results are shown in Fig. 13b. When the separation is small, the interference from a neighbor person is serious, leading to large error ratio of motion counting. However, there are no significant difference between the results of 200*cm* and 250*cm*, which indicates larger separation incurs less interference. It is recommended that a separation distance of 2*m* is appropriate for our configuration.

We study the impact of interference from other persons' motions (even undefined motions). In the evaluation, one volunteer performs squats in the front of a tag while another volunteer perform some irregular movement as interference at a distance of more than 2.5*m* away. We compare the error ratio of counting with and without interference. Fig. 13c shows the results. We can see that even with interference from other person's irregular movement, the error ratio of counting is still less than 6%, which shows that our system works acceptably well even with some interfering activities.

We then examine the error ratio of counting when multiple persons perform same motions simultaneously. In our test, three volunteers perform the same motion. The separation between them is 2*m*. The experimental results are shown in Fig. 13d. The error ratio of counting for different motions are relatively small, about 7% in our test results. Noted that, the error ratio of 'Middle' persons are larger than that of 'Left' and 'Right' person. The reason is that the signal that monitor the middle person contains more interference than that of the boundary ones.

At last, we evaluate the performance when multiple

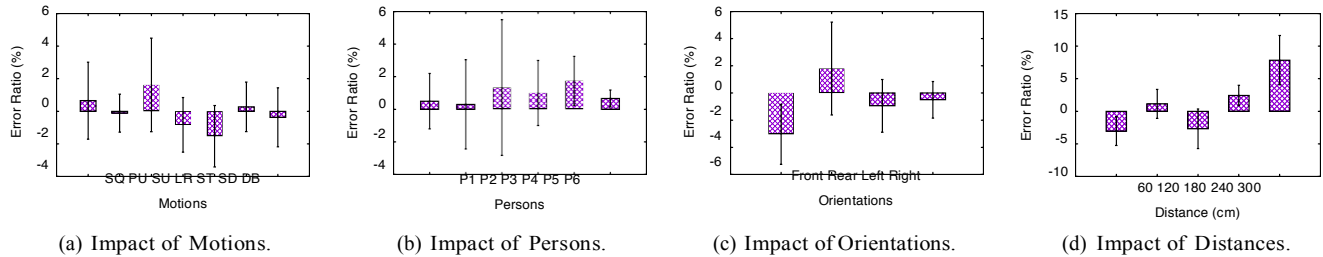

(a) Impact of Motions.  (b) Impact of Persons.  (c) Impact of Orientations.  (d) Impact of Distances.

Fig. 12: Impact on Different Motions, Persons, Orientations and Distances.

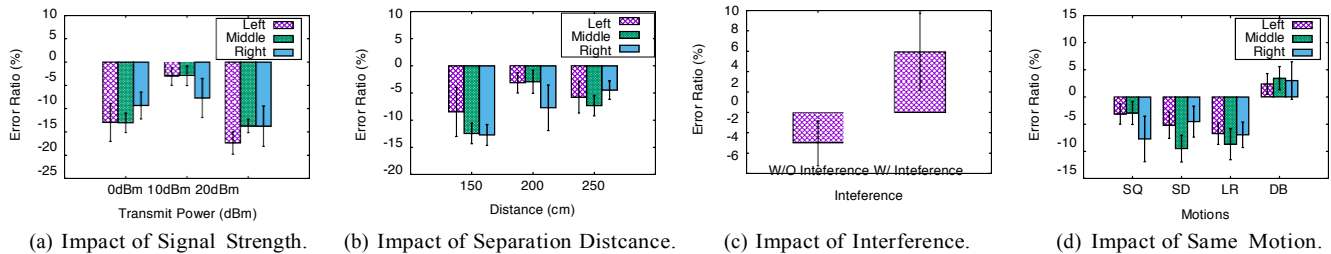

(a) Impact of Signal Strength.  (b) Impact of Separation Distance.  (c) Impact of Interference.  (d) Impact of Same Motion.

Fig. 13: Impact of Multiple Persons.

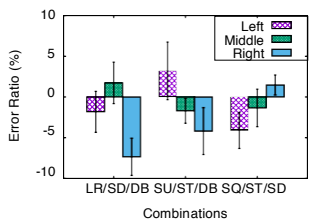

Fig. 14: Impact of Different Motions.

Fig. 15: Confusion Matrix.

TABLE 3: Distribution of Misjudgement Times.

| Misjudgement Times | 0 | 1 | 2 | 3 | Total |
|---|---|---|---|---|---|
| Test Set | 88 | 10 | 2 | 0 | 100 |

persons perform different motions simultaneously. Three volunteers perform different combination of motions. The separation between them is also set as $2m$. As shown in Fig. 14, the tick name of x-axis denotes the combination of motions. For example, 'LR/SD/DB' denotes the 'Left' person performs leg-rise (LR), the 'Middle' person performs stoop-down (SD) and the 'Right' person performs dumbbell (DB). According to the result, it can be seen that the error ratio is relatively small regardless of the combination of motions. Noted that, the error ratio of Dumbbell (DB) is about -7%. The main reason for this is that when people do dumbbell, the cross section of moving upper limbs is small, leading to small variation in received signal, thus making it susceptible to interference from others. It's recommended that people can do motions with obvious limbs movement to achieve better counting accuracy.

**Classification:** We also examine the performance of single-user motion recognition in a given environment. To do this, we collect lots of environmental data and manually divide more than 2500 samples of 7 motions. After that, we extract 10 features from each sample compositing characteristic matrix for cubic SVM model training. The training process only takes a few seconds. We use the 10-fold cross validation to test our model, which reaches 93.1% recognition accuracy. The classification speed of our SVM model is 5000/s. The confusion matrix for classification is shown in Fig. 15, which sums up the result of 10-CV.

Moreover, in order to improve the prediction accuracy, we choose $k = 3$ to make multiple judgments (as described in Sec. 3.5). To verify this result, we select 100 sets of motion data, and only take the first three segments of each group to judge. Our experimental results reach a correct rate of 100%, Table 3 shows the distribution of the number of misjudgments among 100 sets of data. The results show that for continuous repetitive motion, after three or more motions, the classifier can achieve almost error-free recognition.

## 5 RELATED WORK

This section will be divided into two parts, the first part introduces the related work on the passive platform, and the second part is about the relevant work about perception.

## 5.1 Battery-free Backscattering Network

Beginning in 2013, many sessions began to emerge articles about the results of passive platforms. There are many sources of energy in the environment, such as light energy, wind energy, electromagnetic energy, radio (RF) energy [17], [18]. Backscatter based communication received considerable research interests recently, including, ubiquitous energy acquisition, passive protocol optimization and passive component separation. Liu *et al.* [19] use the signal from TV towers to obtain energy for the tag-to-tag communication. Inspiringly, Kellogg *et al.* [20] get the energy from the WiFi APs for backscatter communication. Based on this innovative design, Bharadia *et al.* [9] realize the optimization on conventional WiFi protocol. Kellogg *et al.* [1] propose a revolutionary design, which uses the backscattering signals from the air and keep the ADC component processing. Zhang *et al.* [21] present a low power backscatter system, which allows a tag to embed its information on standard 802.11b codeword to another valid 802.11b codeword. Many solutions have explored the passive projects, One of the important theories they use is frequency shift [22]. In contrast, we study how to use backscattering signals to realize repetitive motion recognition, which leverage the battery-free platform for perception instead of wireless communication.

## 5.2 Device-free Sensing

In recent years, exponential explosive growth of mobile devices once again ignited the people of the new form of human-computer interaction exploration, the use of human-computer interaction to control a wide range of applications [23]. Gesture recognition system, as the basic solution of human-computer interaction, has become more and more popular. The current system allows users to use not only dedicated equipment, but also natural body movements and context-related information. In fact, implantation of gesture recognition systems in electronic products and mobile devices has become commonplace and is on the rise, such as smartphones [24], laptop [25], navigation facility [26] and some game control system [27]. These systems typically implement gesture recognition by utilizing a variety of sensors available on the device, such as computer vision (camera, vidicon, etc.) [27], inertial sensors [28]–[30], vibration sensor [31], acoustics [24], [32], light sensors [33]. However, these technologies still encounter a lot of limitations when they are implemented, such as being customized for specific applications, sensitive to light, high installation costs or high equipment costs, requiring handheld devices, or the need to install additional sensors.

With the proliferation of ubiquitous wireless devices and the establishment of wireless network infrastructure, WiFi-based recognition systems [7], [8], [34], [35]gradually being put forward. WiFi signal can be used not only in gesture recognition, but also localization [8], human identification [6], vibration detection [36] and so on. These WiFi-based systems operate by analyzing the characteristic changes of the wireless signal, such as analyzing changes in CSI (channel state information) or RSSI (received signal strength indication) caused by human motion [37]. Our work leverages the backscattering signals to recognize motions, in order to overcome the limitations in directly applying WiFi signals when multiple users and complicated scenarios are incorporated.

## 6 Conclusion

We present an accurate wireless sensing system building upon a passive backscattering platform. In wireless backscattering systems, human motions could be effectively explored using our customized noise taming and self-adaptive pattern matching algorithms. We evaluate our design with extensive experiments, which shows a satisfying recognition accuracy. Different from previous studies leveraging backscattering technology, our system leverages pervasive wireless APs as RF source instead of customized readers in RFID systems, and enables multiple users peform simultaneously.

### Acknowledgement

This research is partially supported by National key research and development plan 2017YFB0801702, NSFC with No. 61625205, 61632010, 61772546, 61772488, 61520106007, Key Research Program of Frontier Sciences, CAS, No. QYZDY-SSW- JSC002, NSFC with No. NSF ECCS-1247944, and NSF CNS 1526638. NSF of Jiangsu For Distinguished Young Scientist: BK20150030, Tianjin Key Laboratory of Advanced Networking (TANK), School of Computer Science and Technology, Tianjin University. Panlong Yang and Xiang-Yang Li are the corresponding authors.

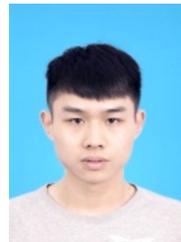

**Ning Xiao** received the B.S. degree in computer science and technology from the College of Computer Science and Technology, University of Science and Technology of China (USTC), China, in 2017. He is now a postgraduate student in USTC. His current research interests include wireless sensing systems, wireless networks.

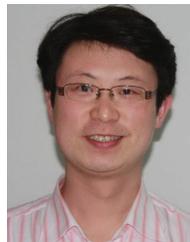

**Panlong Yang** (M'02) received his B.S. degree, M.S. degree, and Ph.D. degree in communication and information system from Nanjing Institute of Communication Engineering, China, in 1999, 2002, and 2005 respectively. During September 2010 to September 2011, he was a visiting scholar in HKUST. Dr. Yang is now a research scientist in School of Computer Science and Technology, University of Science and Technology of China. His research interests include wireless mesh networks, wireless sensor networks and cognitive radio networks. Dr. Yang has published more than 50 papers in peer-reviewed journals and refereed conference proceedings in the areas of mobile ad hoc networks, wireless mesh networks and wireless sensor networks. He has also served as a member of program committees for several international conferences. He is a member of the IEEE Computer Society and ACM SIGMOBILE Society.

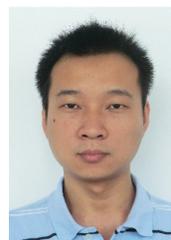

**Yubo Yan** (S'10) received the B.S. degree, M.S. degree, and Ph.D. degree in computer science and technology from the College of Communications Engineering, PLA University of Science and Technology, China, in 2006, 2011 and 2017 respectively. His current research interests include software radio systems, wireless networks, wireless systems. He is a member of the IEEE and the IEEE Computer Society.

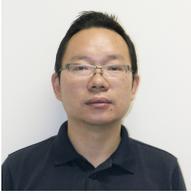

**Hao Zhou** (M'15) received his B.S. and Ph.D. degrees in Computer Science from the University of Science and Technology of China, Hefei, China in 1997 and 2002. He is now an associate professor at the University of Science and Technology of China, and he worked as a project lecturer in the National Institute of Informatics (NII), Japan, from 2014 to 2016. His research interests are in the area of Internet of Things,wireless communication and software engineering.

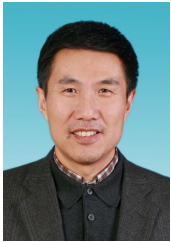

**Xiang-Yang Li** (SM'08) received a Bachelor degree at Department of Computer Science and a Bachelor degree at Department of Business Management from Tsinghua University, P.R. China, both in 1995. He received M.S. (2000) and Ph.D. (2001) degree at Department of Computer Science from University of Illinois at Urbana-Champaign. Xiang-Yang has been an Associate Professor (since 2016) of School of Computer Science and Technology at University of Science and Technology of China, Associate Professor (from 2006 to 2016) and Assistant Professor (from 2000 to 2006) of Computer Science at the Illinois Institute of Technology. He published a monograph "Wireless Ad Hoc and Sensor Networks: Theory and Applications" and co-edited the book "Encyclopedia of Algorithms". His research interests include the cyber physical systems, wireless sensor networks, game theory, and algorithms.